\documentclass[twocolumn, preprintnumbers]{revtex4}

\usepackage{graphicx}         
\usepackage{dcolumn}          
\usepackage{fdsymbol} 
\usepackage{amsfonts}

\usepackage{bm}               
\usepackage{color} 

\newcommand{\rom}[1]{\uppercase\expandafter{\romannumeral #1\relax}}

\newcommand{\supfig}[1]{figure {S#1}}
\newcommand{\supsec}[1]{SM section {\rom{#1}}}

\definecolor{myred}{rgb}{0.8,0,0}

\begin{document}

\title{Cooperative control of environmental extremes by artificial intelligent agents\footnote{This is the arXiv preprint of a current submission. Supplementary Material has not been included.}}

\author{Mart\'i S\'anchez-Fibla$^{1}$, Cl\'ement Moulin-Frier$^{2}$ and Ricard Sol\'e$^{3,4,5}$\footnote{Corresponding author: ricard.sole@upf.edu}}

\affiliation{$^{1}$Department of Information and Communications Technologies, Universitat Pompeu Fabra, 08018, Barcelona, Spain}
\affiliation{$^{2}$Inria - Flowers team Université de Bordeaux ENSTA ParisTech}
\affiliation{$^{3}$Complex Systems  Lab, Universitat Pompeu Fabra, Dr Aiguader 88, 08003   Barcelona, Spain}
\affiliation{$^{4}$ Instituci\'o Catalana de Recerca i Estudis Avançats, Lluís Companys 23, 08010 Barcelona, Spain}
\affiliation{$^{5}$Santa Fe Institute, 1399 Hyde Park Road, Santa Fe NM 87501, USA}

\date{\today}

\begin{abstract}
Humans have been able to tackle biosphere complexities by acting as 
ecosystem engineers, profoundly changing the flows of matter, energy and information. This includes major innovations that allowed to reduce and control the impact of extreme events. Modelling the evolution of such adaptive dynamics can be challenging given the potentially large number of individual and environmental variables involved. This paper shows how to address this problem by using fire as the source of external, bursting and wide fluctuations. Fire propagates on a spatial landscape where a group of agents harvest and exploit trees while avoiding the damaging effects of fire spreading. The agents need to solve a conflict to reach a group-level optimal state: while tree harvesting reduces the propagation of fires, it also reduces the availability of resources provided by trees.  It is shown that the system displays two major evolutionary innovations that end up in an ecological engineering strategy that favours high biomass along with the suppression of large fires. The implications for potential A.I. management of complex ecosystems are discussed.
\end{abstract}

\keywords{Complex systems, cooperation, forest fires, ecosystem engineering, multi-agent reinforcement learning, extreme climate}

\maketitle

 
\section{Introduction}


The term “extreme event” is becoming a common description of a broad class of unanticipated natural events that can have disproportionate social, economic and ecological impacts. Because of the accelerated pace of climate change, megafires, devastating floods and droughts are jeopardizing essential services and infrastructures, from agriculture to biodiversity. These events are expected to become more common in the coming decades \cite{Penuelas2013}. Along with changes in energy use, novel agroforestry practices and conservation policies, intervention scenarios need also to be considered \cite{SoleLevin2022} that take into account the complex, multiscale nature of the problem in space and time \cite{Levin2000,SoleBascomptePUP}. 

The uncertainty associated to environmental fluctuations is far from new to humans. Our ecological success is due to a combination of features that favoured the development of a culturally evolved cooperative social environment \cite{Boyd2009}. In this way, humans became unique in their ways of interacting with the environment. Tool making and social intelligence paved the way for an unprecedented transformation of the biosphere, with humans becoming large-scale {\em ecosystem engineers} \cite{Vitousek1997}  i. e. a species 
having a major impact on the flows of energy and matter \cite{Jones1994}. Agriculture for example can be understood as a powerful way of reducing environmental uncertainty \cite{Gowdy2014}. Similarly, the emergence of urban environments deeply changed our relation with nature and its uncertainties \cite{Maisels2003}. How do these major innovations occur? Moreover, for a given fluctuating environment, what is the role played by learning to find efficient outcomes? 

\begin{figure*}
\begin{center}
\includegraphics[width=18.cm]{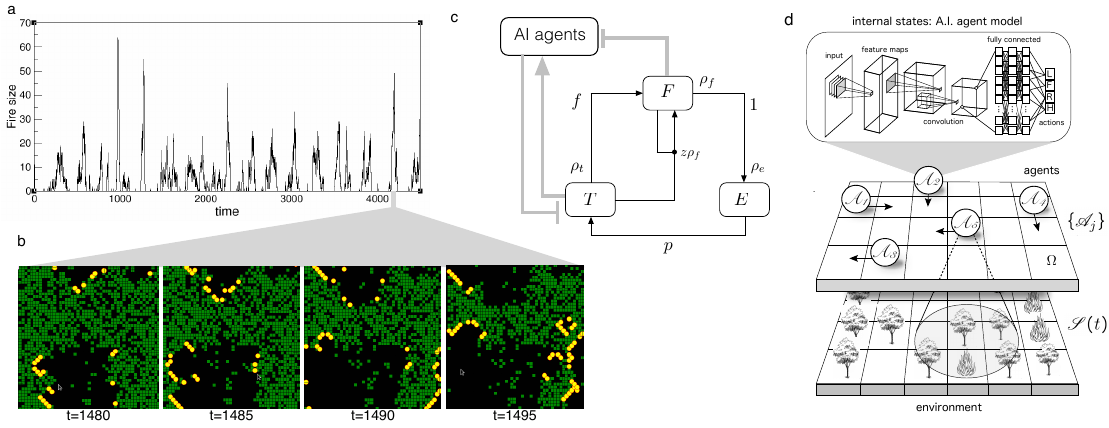}
\caption{\small Forest fire dynamics in time and space. In figure (a) a typical time series of the number of burned sites in a forest fire model (FFM) is displayed for a square lattice with $L=50$ and parameters $p=0.003, f=0.00003$. The number of sites burning (the fire size) shows marked bursting dynamics. Four spatial snapshots are shown in (b) associated with a fire burst. Here, green, yellow and black correspond to trees, fires and ashes (empty sites) respectively. The basic set of rules is summarized in (c) using black arrows. In our model, we add a set of AI agents whose interactions with the environment are marked with grey arrows where positive and negative interactions are indicated as $\rightarrow$ and $\rightfootline$, respectively. They benefit from trees but get punished by fire spreading, and can modify tree density by harvesting trees. In (d) we summarize the levels of interaction between forest fire dynamics and its control by neural agents. The bottom layer defines the observed spatial pattern of states of the Forest Fire Model (FFM), which changes stochastically while can be affected by the action of agents (middle layer) that have a given observation range and can take decisions about their movement and harvesting trees locally. Each agent (upper layer) makes decisions (implements an action policy) by means of a convolutional neural network trained with reinforcement learning. The RL process eventually defines the behavioural pattern displayed by the agent, which translates into a set of potential actions $(LeftTurn,RightTurn,ForwardMove,Harvest)$ in response to the given local complexity of the environment.}
\end{center}
\label{fig:ffm}
\end{figure*}

A central problem, in general, is finding emergent solutions to environmental challenges, such as resource scarcity and the impact of extreme events. Here a long tradition of theoretical models of cooperation, from game theory \cite{Axelrod,Nowak} to statistical physics \cite{Perc2017}, has shown that this kind of behavior emerges under a broad range of conditions. Moreover, the study of future challenges associated to climate change has also benefited from models that include humans and the environment altogether \cite{Anand2022} and allow to understand the potential transitions between dynamical states \cite{Bauch2016}. These models usually consider constant parameters and a description of complex decisions or learning that do not incorporate evolutionary dynamics. Can such evolution between agents and their environment allow finding strategies to deal with environmental extremes? 

In this paper, we approach the previous questions by considering a scenario where a set of artificial intelligent agents interact with a forest where large fires can occur. The agents can learn to exploit a finite set of resources (gathered from trees) while dealing with the destructive potential of forest fires (that require trees to propagate). The dilemma here is quite obvious: a high fraction of tree cover gives larger opportunities for harvesting but also allows for large fires to occur. What kind of strategy can balance this conflict? The case study chosen here allows us to clearly define the constraints imposed by the environment and the repertoire of tasks to be performed by the agents. The choice of fire is grounded in its ecological impact on a wide range of ecosystems all over the planet and its role in human history \cite{Goudsblom1992,Bond2005}. Fire burns ecosystems acting as a major evolutionary force. At the same time, its use by humans is connected to a major innovation used by our species as a way of engineering the wild. Here we approach the problem by using a hybrid model that combines forest fire spreading with a set of artificial learning agents that adapt to environmental conditions and exploit resources while finding ways to control fire. The outcome of the evolution of these agents is the emergent cooperative management of flammable ecosystems that provides both a higher tree yield together with a marked fire reduction.

Agents will evolve as they interact and modify fire behaviour using reinforcement learning (RL) \cite{Sutton1998}. RL is one of the key directions of Machine Learning and is playing a major role in ongoing developments within Artificial Intelligence (AI, see \cite{Russell2016}), spanning multiple domains, from game theory to advanced robotic tasks. In a nutshell, in RL, optimal strategies emerge within the lifetime of an agent (both learner and decision-maker) who interacts with an environment, giving rise to (cumulative) rewards that the agent learns to maximise. Since this is an unsupervised process, RL provides a powerful approach to exploring social dilemmas \cite{Perolat2017} or the learning and maintenance of social norms within societies \cite{Koster2022}. As shown below, cooperative strategies emerge as agents deal with fire, evolving different forms of exploiting resources while protecting themselves from damage,  first using a simple clustering strategy and later on developing a more sophisticated one where decision-making depends on a distributed management of the local environment.

 


Modelling the evolution of ecological control by a set of agents requires two main components. The first is a model of extreme environmental fluctuations provided by fire spread. The second is the formalization of a set of agents that must respond to these fluctuations while gathering resources from trees. The first part is done using a minimalist framework based on cellular automata (CA), which have been successfully used in many areas \cite{Chopard2000,Ilachinski2001,Batty2007}. The second involves a set of agents that learn to control behavior through decentralized Multiagent Reinforcement Learning (MARL) \cite{Perolat2017,Koster2022}. Each agent learns its own policy through Deep Reinforcement Learning (DRL) \cite{Mnih2015} without having access to the observations, actions and rewards of the others.


{\bf Spatial dynamics of fire}. Fire spread is described by means of the {\em Forest Fire Model} \cite{Bak1990,Drossel1992} where a toy description of fire spreading is applied to the states of each site on a two-dimensional, $L \times L$ lattice $\Omega$. For convenience we use periodic boundary conditions (i. e. dynamics takes place on a torus). Each site ${\bf r} \in \Omega \in \mathbb{Z}^2$ can be in three possible states, namely $S({\bf r}) \in \Sigma=\{E,T,F\}$, where $E$ denotes an empty cell, $T$ a tree cell and $F$ a fire cell. The state of the system, ${\cal S}$ will be updated by means of three probabilistic events, namely: (1) spontaneous burning of a tree, i. e. a transition $$T \buildrel f \over \longrightarrow F$$ at a rate $f$, leading to a burning site (fire cell); (2) growth of new trees from empty sites, i. e. with a probability $p$ we have $$E \buildrel p \over \longrightarrow T.$$ 
(3) The last rule allows fire propagation: if a given tree has a neighbour that is a fire, it burns too. This means a (deterministic) transition 
$$T \buildrel 1 \over \longrightarrow F.$$ Hereafter, the set of neighbours $\Gamma({\bf r})$ is defined by a {\it von Neumann neighborhood}, i. e. the four nearest ones. For our two-dimensional lattice, we have ${\bf r}=(i,j)$ and $\Gamma({\bf r})= \{ (i\pm 1,j), (i, j\pm 1) \}$.

It can be shown (see \supsec{1}) that the previous discrete rules allow to define a mathematical model of forest fire dynamics that converges to a stable attractor (fixed point) where fires and trees coexist. However, for $p,f \ll 1$ and $f \ll p$ the actual discrete, spatially-explicit dynamics is highly fluctuating, exhibiting a broad spectrum of fluctuations (a self-organized critical (SOC) state) that includes extreme events  \cite{Bak1996,Turcotte1999}. In figure 1a-b we show an example of the fire spreading dynamics on a $L=50$ lattice for $p=3 \times 10^{-3}$ and $f=3 \times 10^{-5}$ with four snapshots (b) associated to a major fire event. The origins of such extreme events are to be found in the separation of time scales associated to the SOC dynamics \cite{Drossel1992,Dickman2000}. Despite of its simplicity, the FFM and variations of it has been successfully applied to model the statistical patterns of the actual fires \cite{Malamud1998,Pueyo2007,Pueyo2010,Hantson2015,VanNes2018}. Other similar models that exhibit SOC have been used to study other systems displaying wide fluctuations, including earthquakes, rainfall or financial markets \cite{Bak1996,Jensen1998}. 

The basic rules and transitions are summarized in figure $1c$, along with the schematic interaction with agents, which benefit from the presence of trees but need to avoid the damage caused by fires. Because of the fluctuating nature of fires (which can exhibit large peaks of destruction, see figure $1b$), resource availability and potential damage by fire can be rather unpredictable. 
It can be shown (see \supsec{1}) that the previous discrete rules allow to define a mathematical model of forest fire dynamics that converges to a stable attractor (fixed point) where fires and trees coexist. However, for $p,f \ll 1$ and $f \ll p$ the actual discrete, spatially-explicit dynamics is highly fluctuating, exhibiting a broad spectrum of fluctuations that includes extreme events  \cite{Bak1996,Turcotte1999}. In figure 1a-b
we show an example of the fire spreading dynamics on a $L=50$ lattice for $p=3 \times 10^{-3}$ and $f=3 \times 10^{-5}$ with four snapshots (b) associated to a major fire event. Trees, fires and empty sites are indicated as green, yellow and black colors, respectively. The origins of such extreme events are to be found in the separation of time scales associated to the SOC dynamics \cite{Drossel1992,Dickman2000}. Despite of its simplicity, the FFM and variations of it has been successfully applied to model the statistical patterns of the actual fires \cite{Malamud1998,Pueyo2007,Pueyo2010,Hantson2015,VanNes2018}. Other similar models that exhibit SOC have been used to study other systems displaying wide fluctuations, including earthquakes, rainfall or financial markets \cite{Bak1996,Jensen1998}.


{\bf Agent-environment interaction}. We consider a population of $N$ learning agents $\{{\cal A}_j \}$ with $j=1,...,N$, interacting with the environment as defined by ${\cal S}(t)$. Initially ($t=0$), each agent is placed randomly on $\Omega$ and given a random orientation (either North, East, South or West). Each cell of the grid-world environment can contain one agent (fig. 1d) which can influence the state of ${\cal S}(t)$ by executing some actions, as defined below. In order to describe the profit tied to the exploitation of trees, we introduce a resource associated to a tree that can be a source of reward for the learning agent. Specifically, a tree can carry a resource (say a fruit) that can be consumed by agents. Let us indicate as $T^*$ these fruit-carrying tree. Once consumed, it can be restored after some time, given by a recovery rate. This extra state does not modify the fire dynamics, since it does not affect the FFM rules. 

At each time step, each agent can randomly choose to move forward ($a_F$) to a neighbouring site if not occupied by another agent, rotate to the left ($a_L$), rotate to the right ($a_R$), or harvest the site in front of it ($a_H$). If, after moving, the new site is a resource it will be consumed by the agent, i.e it will become a tree cell (which will be able to regenerate a resource with probability $P_r$).  The harvesting $a_H$ action will only have an effect if the cell next to the agent in its front direction is occupied by a tree with no available resources. In that case, the tree will be removed (i. e. replaced by an empty cell). An agent consuming a resource will receive a positive reward $R_r$, while an agent residing on a fire cell will receive a negative reward $R_f$ (with $R_r$ much smaller than $R_f$). We can summarize these extra transitions 
as follows: 
$$T^* \buildrel {\cal A} \over \longrightarrow T \buildrel P_r \over \longrightarrow T^*$$
where the first transition indicates that the presence of an agent implies the loss of the resource. 


\begin{figure*}[t!]
	\begin{center}
		\includegraphics[height=11 cm]{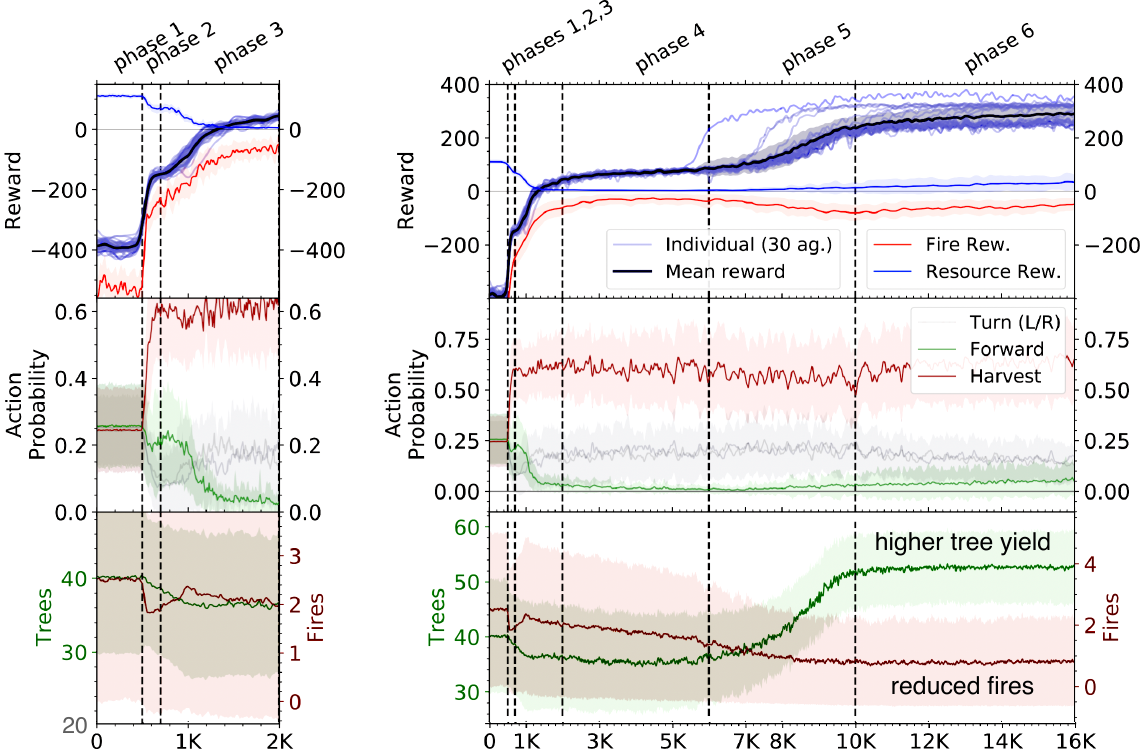}
		\caption{Evolving cooperation by artificial intelligent agents. The time series of all the relevant measures of a population of RL agents that evolve ecological engineering strategies is shown over $16K$ episodes. This includes the reward for all agents, mean action probability usage of all agents, number of trees (mean and standard deviation), number of fires (mean and standard deviation). The left column is a zoomed view of the first $2K$ episodes, with $700$ time steps each. Parameters are: $p_t=0.08, p_f=0.005$ for the FFM and $L=10, N=30$ with a $7 \times 7$ observation window. On the right the full time series is displayed.  Notice the marked increase of tree yield and the suppression of fires, which also involves a drastic reduction of fluctuations.}
		\label{fig:learning}
	\end{center}
\end{figure*} 


{\bf Agent learning}. The task of each agent is to maximize its own reward, i.e. to maximize the number of collected resources while avoiding to be burned by fire. Learning is structured in a sequence of episodes, with a fixed duration of $T$ time steps each. At the start of each episode, a new map is randomly initialized with tree and empty cells according to a probability distribution $p_{init}$ and $N$ agents are randomly positioned on the map (random positions and orientations). 

Each agent learns its own action policy, mapping its partial observation of the environment at the current time step to a probability distribution over its actions. 
The learning objective is to maximize a cumulative reward obtained over an entire episode:
\begin{equation}
G_t = \sum_{t=0}^{T-1} \gamma^t R_{t} 
\end{equation}
where $R_t$ is the reward obtained by the agent at time step $t$ and $\gamma < 1$ is a discount factor.

The action policy described above is generated by a 2D Convolutional Neural Network in which every agent trains independently in order to maximize its cumulative reward $G_t$ (details are given in \supsec{3} of the supplementary material). The network weights are updated after every experience (state, action, next state and reward). The agents act randomly at the start given the random initialization of weights. During training they are able to maintain a certain level of exploration by favoring (with a small contribution) the training loss towards the equiprobable distribution of actions.

\begin{figure*}
	\begin{center}
		\includegraphics[width=15 cm]{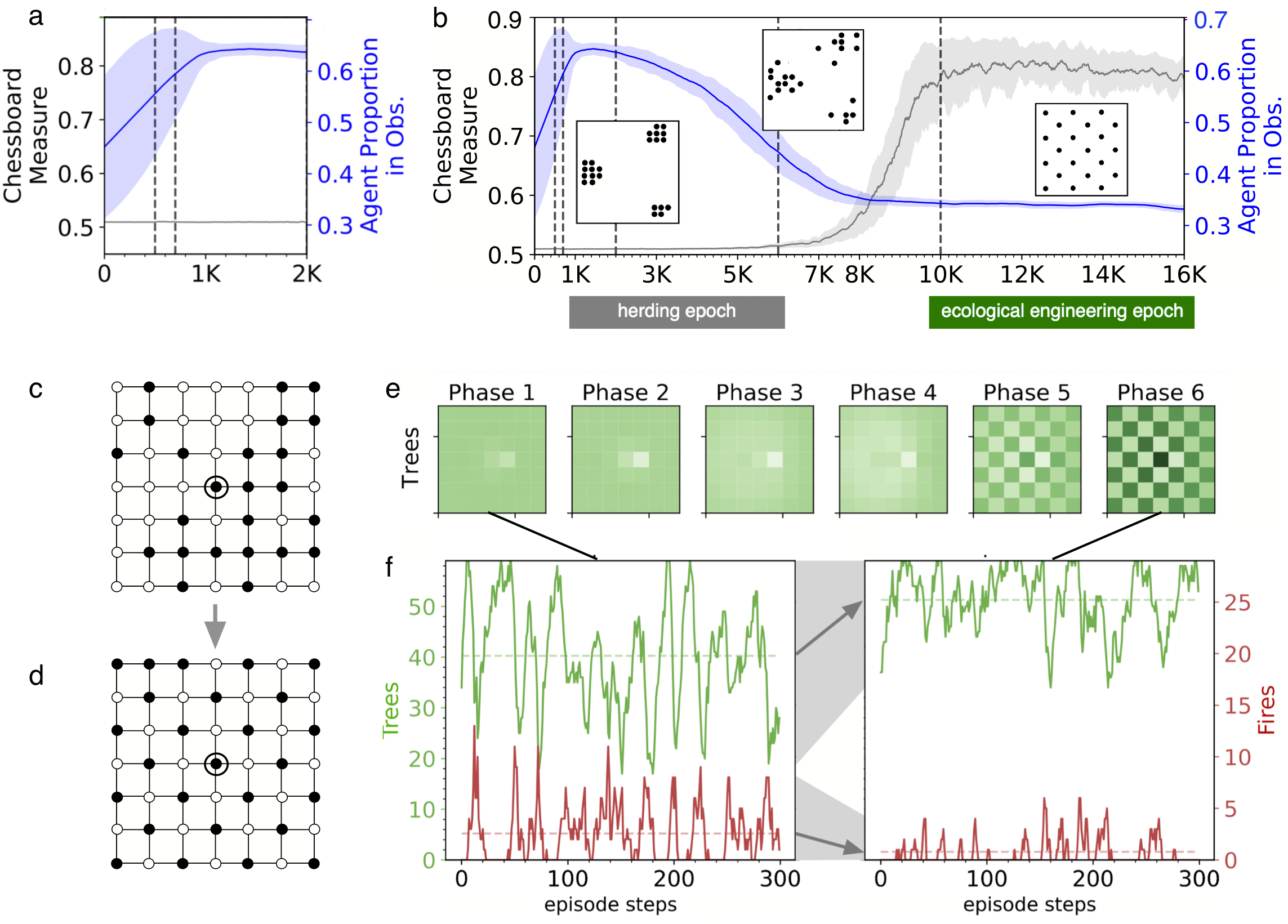}
        
        \caption{Agent perception and its impact on FF dynamics. In (a-b) the time series for the chessboard measure ($\cal C$) is displayed. The three insets in the bottom right figure sketch the three spatial (average) arrangements of agents associated to the herding and ecological engineering epochs, respectively. The RL system evolves from an initial state dominated by fire dynamics to an agent-controlled environment where average observations indicate a well-defined chessboard pattern with trees located (on average) in a regular fashion. The chessboard measure is obtained by averaging the agent observations ($O^{\text{Tree}}_{a,t}$) for all agents on the $7 \times 7$ domain (c-d), involving the presence (filled circle, $1$) or absence (empty circle, $0$) at a given time. The (c) and (d) lattices sketch initial and final states. The open circle indicates the location of the agent $\cal A$.  In (e) we show the measured mean densities $M({\bf r})$ from all agent observations at the end of each phase. Notice the development of a well-defined chessboard. The heat maps have the same shape as agent's observations (i.e. 7x7, centered on the agent), with the agent's front direction pointing to the right. The initial and final  time series (f) within one episode and shown. Notice the reduction of the variance of fluctuations in both tree and fire time series, and the marked improvement in tree yield along with fire suppression.}
		\label{fig:learning_obs}
	\end{center}
\end{figure*}

 
\section{Results}


Under the environmental description provided by the system's state $\cal S$, the agents benefit from tree-rich neighbourhoods, since trees provide the resources and reward. But trees also propagate fire and agents are highly punished when being burned (through a negative reward), thus creating an incentive to harvest the trees around them. The agents must therefore learn a strategy that maximizes the number of trees around them while minimizing fire propagation. How will the agents solve this conflict? As shown below, several rapid changes in agent behavior take place, involving two marked cooperative transitions. The model reveals several phases of evolution where agents learn to protect themselves from fire first by clustering in groups (the herding epoch) while, on the long run, ecological engineering is developed as a spatially-distributed decision making pattern (the ecological engineering epoch).

Figure~\ref{fig:learning} (left) shows the six phases which spontaneously emerge from the interaction between fire spreading and learning dynamics.
Training starts at the end of this interval, and a rapid increase in harvesting (and a decrease in the rotation actions) is observable, while maintaining the forward action approximately in order to collect resources. The mean reward raises consequently, yet is still highly negative which indicates that foraging and harvesting are not performed efficiently (fig.2, left). Shortly afterwards, a cooperative transition starts to emerge. In order to characterize it, some quantitative definitions are needed (see below). Figures \ref{fig:learning} and \ref{fig:learning_obs} correspond to a simulation of 16k episodes of $T=500$ steps each with $N=30$ agents in a $10 \times 10$ grid and the FFM model having fixed probabilities of tree regrowth $p_{tree} = 0.08$ and fire appearance $p_{fire} = 0.008$. The results presented here are consistent when using larger lattices (see SM).
As a reference state, we first run our simulations without learning. Throughout $500$ episodes, random choices occur and low rewards are observed (as expected). In this phase, the number of trees and fires are consistent with the FFM predictions (see \supsec{1}).

Consider $O_{a,t,e}$ to be the observation matrix (containing all perceptual channels: trees, fires, resource and agents) of agent $a$ at time step $t$ of episode $e$ centered on the agent. For instance, it will have $O^{\text{Tree}}_{a,t,e}=1$ value if a tree is observed and zero otherwise. We then define the matrices corresponding to the mean observations of all the agents for every channel $c \in \{ \text{Tree}, \text{Fire}, \text{Resource}, \text{Agent} \}$. 
\begin{equation}
M^{c}_e(i,j) = \frac{2}{T} \sum_{t=250}^{T=500} \left( \frac{1}{N} \sum_{a=1}^{N} O^c_{a,t,e}(i,j) \right)  
\label{eq:means}
\end{equation}
For instance, $0 \le M^{\text{Tree}}_e \le 1$ is the normalized mean of all the tree observation matrices $O^{\text{Tree}}_{a,t,e}$ (the tree channel of the observation) for all agents ($a$) and the final time steps ($T$) of the episode $e$ of length $T$, and similarly for all other channels. If no episode is indicated as subscript when referencing $M$, the matrix is assumed to be computed also for the mean of the last episodes of the learning ($E=16K$ in figure 2):
\begin{equation}
M^{c}(i,j) = \frac{2}{E} \sum_{e=E/2}^{E} M^c_{e}(i,j)  
\label{eq:means2}
\end{equation}
In figure \ref{fig:learning_obs} we plot the mean tree densities corresponding to the $M^{\text{Tree}}_e$ matrices for the different phases of the learning. In the supplementary material we plot $M^{c}$ matrices for all channels $c$ in \supfig{2}.

In order to characterize the spatial arrangement of agents, we also introduce the clustering ratio per episode, as measured by the local average: 
\begin{equation}
d(e) = {1 \over 8} \sum_{i,j \in \Delta({\bf r})} M^{\text{Agent}}_{a,e}(i,j)   
\end{equation}
with $\Delta({\bf r})= \{ (i, j\pm 1), (i\pm 1, j), (i\pm 1, j\pm 1) \}$ being the eight cells in the Moore neighborhood. This is nothing but the probability that, given an agent $\cal A$ located at $\bf r$, there are other agents in its nearest neighbourhood. 

{\bf Clustering phase}. Using the clustering measure, we can detect a first cooperative change, which we label the {\it herding epoch}. It is characterized by high values of agent packing, constantly increasing and reaching a maximum from episodes $1000$ to $2000$ as seen in figure \ref{fig:learning_obs}b. A sudden drop in the forward action takes place along with an increase in the rotation actions, while harvesting occurs at a high rate (around $60$\% of the time). These measures indicate that the agents learn to rapidly form a packed group at the beginning of the episode, then stay in place while harvesting trees around them, creating a safe area where they are efficiently protected from fire. As a consequence, the proportion of fire cells decreases (but is still relatively high due to fire propagation outside of the group area). We observe that the negative rewards received when agents are burned by fire and the positive rewards from collected resources both approach $0$, indicating that the agents are well protected from fire but are not able to collect resources within the packed group. While sub-optimal, this herding strategy still makes sense at this stage since the penalty for getting burned $R_f$ is much stronger than the positive reward $R_r$ of consuming a resource.

{\bf Expansion phase}. In the next phase in the evolution of agent behaviour (episodes 2000 to 6000), the agents improve upon the packing strategy discovered in the previous phase. While reward grows (figure 2) they maintain the group coherence while allowing more space between agents within the group, as indicated by the increase in the agent distance measure: clusters start to expand in space, occupying larger areas. The proportion of fire cells in the environment continues to decrease, since the area of the grid covered by the group increases and the spaces newly introduced between the agents are too small for fire to propagate. These observations indicate that a new collective strategy is building up. The aftermath of the previous phase (episodes 7000 to 10000, phase 5) is marked by an increase in reward while agents become more isolated, along with a reduction of fires and an increase in tree cover. This phase defines the transient towards a new phase characterized by the dominance over fluctuations and the development of fire suppression. How is this achieved? This transition is the result of a spatially-extended control of fire spread resulting from a new set of decisions based on a more accurate control of the agents over their environment. This can be quantified by means of a global {\it chessboard} measure $\cal C$. The motivation of this measure is rooted in the emergence of a behavioral pattern where active harvesting of most close four neighboring trees occurs, while those in the four diagonals (where fire cannot propagate) tend to be free from harvesting.

As defined above in equations \ref{eq:means} and \ref{eq:means2}, $0 \le M^{\text{Tree}} \le 1$ is the normalized mean of all the tree observation matrices $O^{\text{Tree}}$ (the tree channel of the observation) for all agents ($a$), second half of the steps of the final episodes.  The chessboard measure $\cal C$ captures in a single number how close to a chessboard pattern is the system's state. It is in fact a local measure computed from the normalized agents observation mean tree matrices $M^{\text{Tree}}$. We can define $\cal C$ as:
\begin{equation}
{\cal C} = {1 \over Z} 
\sum_{{\cal A}({\bf r})} \left ( 
\sum_{{\bf r}'\in \Gamma_1(\circ)} 
 (1-M^c({\bf r}'))
+ \sum_{{\bf r}'\in \Gamma_2(\bullet)}
 M^c({\bf r}') \right )
\end{equation}
where $Z=2L^2qN$ is a normalization factor and $M^{\text{Tree}}$ the mean of the tree observations (thus capturing the mean number of trees seen around every agent). If ${\bf r}=(i,j)$, we define the local subsets $\Gamma_1(\circ)$ and $\Gamma_2(\bullet)$ as they are defined by the lattice of figure \ref{fig:learning_obs}d. As defined, it can be checked that ${\cal C}=0.5$ for both empty and fully occupied lattices while ${\cal C}=1$ for a full chessboard pattern.

\begin{figure*}
	\begin{center}
		\includegraphics[width=18cm]{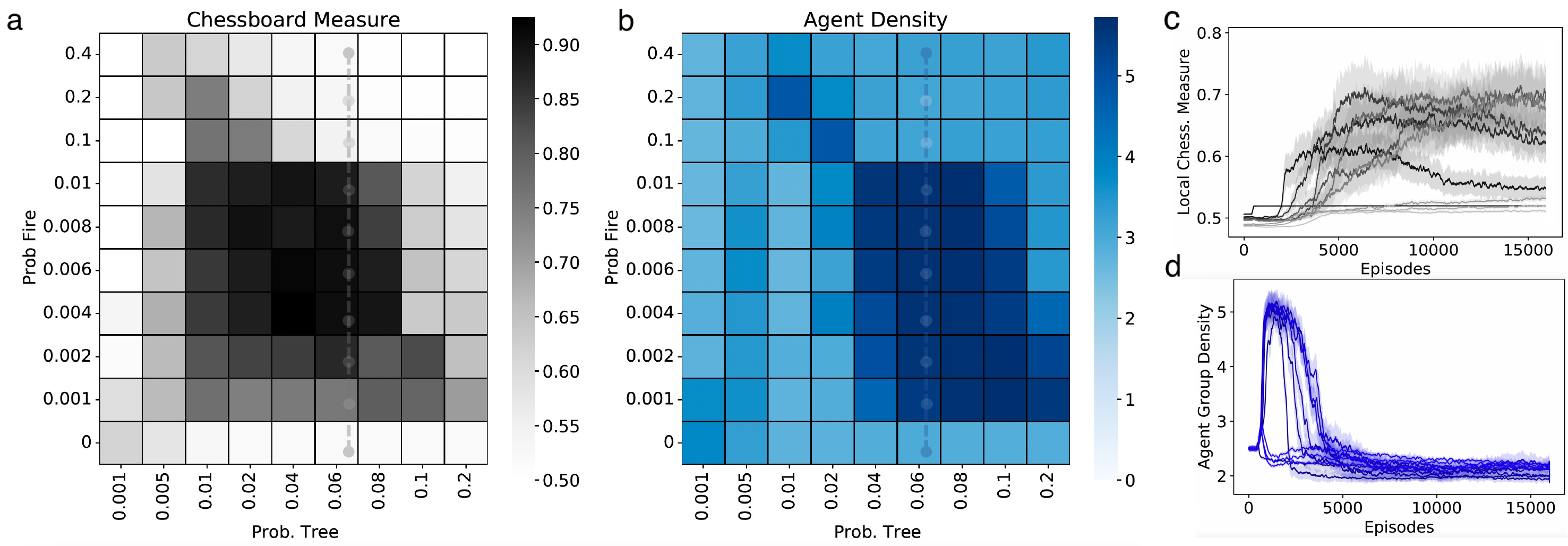}
		\caption{The cooperation phase space. In (a), the maximum chessboard measure is shown as a function of tree regrowth and spontaneous fire appearance. Cooperation is maximum (with a chessboard measure of $0.92$) at $p_{tree} = 0.04$ and $p_{fire} = 0.004$ and collapses with increasing probability of tree growth ($p_{tree} = 0.1$). In (b) the maximum agent density measure is displayed. The evolution per episode of the chessboard measure of 10 simulations is shown in (c) (with 16K episodes each), with all the simulations corresponding to the ten values of the column with $p_{tree} = 0.06$ (as indicated in the grid by the dashed dotted line).  In (d) the corresponding evolution per episode of the agent density measure for the same parameters and 10 simulations (column $p_{tree} = 0.06$) of figure (c).}
		\label{fig:chessboard}
	\end{center}
\end{figure*}

In this fifth phase, we observe a sudden increase in the chessboard measure, indicating that the agents learn to harvest trees in a much more structured way. As mentioned at the beginning of this section, the chessboard pattern is predicted as an optimal structure that can prevent fire propagation while allowing trees (and therefore resources) to appear on half of the grid cells. The effectiveness of this pattern is confirmed by a substantial increase in the proportion of trees in the grid, even though the harvest action continues to be executed at the same rate as in the previous phase. We also observe that the proportion of fire cells continues to decrease, confirming the effectiveness of the chessboard pattern in reducing fire propagation, thus increasing the mean reward. 

{\bf Ecological engineering phase}. The last phase in the evolution of our RL agents (Episodes 10k to 16k) involves control over fire spread. We label it the {\it ecosystem engineering epoch} (marked with a colour bar in figure 3b). The chessboard measure remains constant at its maximum value in this sixth phase. The agents benefit from this well-engineered ecosystem, as shown by the convergence of the mean reward towards its maximum. However, we observe a significant variance in the reward, with a few agents obtaining much more reward than others. However, the higher reward received by these agents does not seem to negatively impact the reward of others, suggesting that they are more "risk-takers" than "free-riders", i.e. agents that take the risk of moving across the grid to collect more rewards without dramatically impacting the structure of the global chessboard pattern.

In figure~\ref{fig:chessboard} we take a deeper look into the emergence of the two identified cooperative strategies: agent grouping and chessboard arrangement (through the agent density and chessboard measures respectively). We study the robustness of cooperation for different environmental conditions, considering different probabilities of tree regrowth $p_{tree}$ (ranging from $10^{-3}$ to $10^{-1}$) and fire appearance $p_{fire}$ (ranging from 0 to 0.4). Each cell of the grid (figure~\ref{fig:chessboard}a-b)  corresponds to a simulation of 16k episodes and the maximum of the chessboard measure (figure~\ref{fig:chessboard}a) and the maximum of the agent density measure (figure~\ref{fig:chessboard}b) are plotted. We can observe that cooperation is maximum at particular regions of this space and collapses for particular conditions. For instance, the highest level of cooperation and environment engineering happens at the probability of tree regrowth $p_{tree}=0.06$ (and fire appearance probability $p_{fire}=0.006$) but collapses just after that (at $p_{tree}=0.1$). When the rate of tree growth is too high the environment becomes too challenging to be engineered as a group, with fire rapidly propagating, and individualistic strategies are more beneficial. This collapse of cooperation happens as well in the rest of the extremes: when there is no fire, when there is too much fire and when there are too few trees (see also the SM for a full hyper-parameter analysis: \supfig{6} and \supfig{7}).


\section{Discussion}

 
In this paper we have shown how modelling learning dynamics is an important step towards understanding how can a collective of cognitive agents achieve cooperative control over extreme events. This is nowadays a timely issue, as global warming is rapidly disrupting the long temperature stability that allowed human civilization to thrive during the Holocene \cite{Ellis2015}. 

Human societies have been particularly successful in this respect by developing cooperation strategies \cite{Carballo2013}. Cooperation is thus a major force of nature control. But how such a control is achieved when conflicting constraints arise? Previous work has explored the dynamics of cooperation under game-theory approximations \cite{Axelrod,Roca2009} including those experiencing noisy conditions \cite{Helbing2009}. In other studies, the use of neural agents allowed to explore the emergence of simple cooperative strategies \cite{Burtsev2006} and the interplay between cooperation and social intelligence \cite{McNally2012}. However, our study is, to the best of our knowledge, the first work that considers the rise of cooperation among cognitive (RL) agents dealing with extreme events. When dealing with wide fluctuations and conflicting constraints as those addressed here, the agents must develop novel, cooperative strategies. 

We have shown that the outcome of these conflicts are several consecutive transitions that provide increasing opportunities to the agents, including two major events that reflect the partial and eventually global control of the entire ecosystem. This example illustrates the potential for A.I. systems to help exploring novel ways to deal with the high-dimensional nature of complex environments. By suppressing fires, agents have effectively taken control of uncertainties, while also obtained stable resources with high yield. This illustrates the emergence of ecosystem engineering on a global scale. Moreover, our work shows how RL models can help to explore other human-ecological transitions under a synthetic approximation \cite{Sole2016}. Finally, although ours is a simple model, we believe that it illustrates the potential that A.I. systems can have to find solutions to the management of other complex ecosystems facing tipping points or being damaged by extreme events.

\begin{acknowledgments}
\small{The authors thank the Complex Systems Lab members for fruitful discussions. Special thanks to Hari Seldon for his inspiring ideas. RS was supported by the Spanish Ministry of Economy and Competitiveness grant FIS2016-77447-R MINECO/AEI/FEDER, and the Santa Fe Institute. CMF was supported by the Inria Exploratory action ORIGINS (\url{https://www.inria.fr/en/origins}) as well as the French National Research Agency (\url{https://anr.fr/}, project ECOCURL, Grant ANR-20-CE23-0006). MSF was supported by the Spanish Ministry of Economy and Competitiveness grant DPI2016-80116-P MINECO/AEI/FEDER. }
\end{acknowledgments}


\begin{references}

\bibitem{Mnih2015}
Mnih, V., Koray K., Silver D., A. Rusu, Veness J., G. Bellemare,  Graves A. et al. 2015. Human-level control through deep reinforcement learning. Nature 518, no. 7540: 529-533.

\bibitem{Penuelas2013}
Pe\~nuelas, J., Sardans, J., Estiarte, M. et al., 2013. Evidence of current impact of climate change on life: a walk from genes to the biosphere. Global change biology 19, 2303-2338.

\bibitem{SoleLevin2022}
Sol\'e, R. and Levin, S., 2022. Ecological complexity and the biosphere: the next 30 years. Phil. Trans. Royal Soc. B 377, 20210376.


 \bibitem{Levin2000}
 Levin, S.A., 2000. Multiple scales and the maintenance of biodiversity. Ecosystems 3, 498-506.
 
\bibitem{SoleBascomptePUP}
Sol\'e, R. and Bascompte, J. 2007. {\it Self-organization in complex ecosystems}. Princeton U. Press. Princeton, USA.


 \bibitem{Boyd2009}
 Boyd, R. and Richerson, P.J., 2009. Culture and the evolution of human cooperation. Phil. Trans. R. Soc. B 364, 3281-3288.


\bibitem{Vitousek1997}
Vitousek PM et al (1997) Human domination of Earth ecosystems. Science 27: 494-499.


\bibitem{Jones1994}
Jones CG, Lawton JCG and Shachak M. 1994. Organisms as ecosystem engineers. Oikos 69: 373-386.
	
\bibitem{Gowdy2014}
Gowdy, J. and Krall, L., 2014. Agriculture as a major evolutionary transition to human ultrasociality. Journal of Bioeconomics, 16, 179-202.
	
\bibitem{Maisels2003}
Maisels, C.K., 2003. {\it The emergence of civilization: From hunting and gathering to agriculture, cities, and the state of the near east}. Routledge.

\bibitem{Axelrod}
Axelrod, R. 2006. {\it The evolution of cooperation}. Basic Books, New York. 

\bibitem{Nowak}
Nowak, M.A., 2006. Five rules for the evolution of cooperation. science, 314(5805), pp.1560-1563.
	
\bibitem{Perc2017}
Perc, M., Jordan, J.J., Rand, D.G., Wang, Z., Boccaletti, S. and Szolnoki, A., 2017. Statistical physics of human cooperation. Phys. Rep. 687, 1-51.

\bibitem{Anand2022}
Farahbakhsh, I., Bauch, C.T. and Anand, M., 2022. Modelling coupled human–environment complexity for the future of the biosphere: strengths, gaps and promising directions. Phil. Trans. R. Soc. B 377(1857), 20210382.

\bibitem{Bauch2016}
Bauch, C.T., Sigdel, R., Pharaon, J. and Anand, M., 2016. Early warning signals of regime shifts in coupled human–environment systems. Proc. Natl. Acad. Sci. USA 113, 14560-14567.

\bibitem{Goudsblom1992}
Goudsblom, J., 1992. The civilizing process and the domestication of fire. Journal of World History 3, 1-12.

\bibitem{Bond2005}
Bond, W.J. and Keeley, J.E., 2005. Fire as a global ‘herbivore’: the ecology and evolution of flammable ecosystems. Trends Ecol. Evol. 20, 387-394.
	
\bibitem{Sutton1998}
R. Sutton, A. Barto, {\em Reinforcement learning: An introduction}, Cambridge, MIT press, 1998.
	
\bibitem{Russell2016} 
Russell, S.J. and Norvig, P., 2016. {\em Artificial intelligence: a modern approach}. Prentice and Hall, New Jersey.	
	
\bibitem{Perolat2017}
Perolat, J., Leibo, J.Z., Zambaldi, V., Beattie, C., Tuyls, K. and Graepel, T., 2017. A multi-agent reinforcement learning model of common-pool resource appropriation. Advances in neural information processing systems, 30.

\bibitem{Koster2022}
Köster, R., Hadfield-Menell, D., Everett, R., Weidinger, L., Hadfield, G.K. and Leibo, J.Z., 2022. Spurious normativity enhances learning of compliance and enforcement behavior in artificial agents. Proc. Natl. Acad. Sci. USA 119, e2106028118.

\bibitem{Chopard2000}
Chopard, B. and Droz, M., 2000. {\it Cellular Automata Modelling of Physical Systems}.Cambridge University Press.

\bibitem{Ilachinski2001}
Ilachinski, A., 2001. {\it Cellular automata: a discrete universe}. World Scientific Publishing Company.
 	
\bibitem{Batty2007}
Batty, M., 2007. {\it Cities and complexity: understanding cities with cellular automata, agent-based models, and fractals}. The MIT press.

\bibitem{Bak1990}
Bak, P., Chen, K. and Tang, C., 1990. A forest-fire model and some thoughts on turbulence. Physics letters A, 147(5-6), pp.297-300.

\bibitem{Drossel1992}
Dr\"ossel, B. and Schwabl, F., 1992. Self-organized critical forest-fire model. Physical review letters 69, 1629.

\bibitem{Bak1996}
 Bak P (1996) {\it How nature works. The science of self-organised criticality}. Copernicus, New York
	
\bibitem{Turcotte1999}
Turcotte, D.L., 1999. Self-organized criticality. Reports on progress in physics 62, 1377.
	
\bibitem{Dickman2000}
Dickman, R., Muñoz, M.A., Vespignani, A. and Zapperi, S., 2000. Paths to self-organized criticality. Brazilian Journal of Physics 30, 27-41.

	
\bibitem{Malamud1998}	
Malamud, B.D., Morein, G. and Turcotte, D.L., 1998. Forest fires: an example of self-organized critical behavior. Science 281, 1840-1842.

\bibitem{Pueyo2007} 	
Pueyo, S., 2007. Self-organised criticality and the response of wildland fires to climate change. Climatic Change 82, 131-161.	
	
\bibitem{Pueyo2010} 
Pueyo, S., De Alencastro Graça, P.M.L., Barbosa, R.I. eta l., 2010. Testing for criticality in ecosystem dynamics: the case of Amazonian rainforest and savanna fire. Ecology letters 13, 793-802.

\bibitem{VanNes2018}
Van Nes, E.H., Staal, A., Hantson, S. et al., 2018. Fire forbids fifty-fifty forest. PloS one 13, e0191027.
	
\bibitem{Hantson2015}
Hantson, S., Pueyo, S. and Chuvieco, E., 2015. Global fire size distribution is driven by human impact and climate. Global Ecology and Biogeography, 24, 77-86.

\bibitem{Jensen1998}
Jensen, H.J., 1998. {\it Self-organized criticality: emergent complex behavior in physical and biological systems}. Cambridge U. Press. Cambridge UK.

\bibitem{Ellis2015}
Ellis, E.C., 2015. Ecology in an anthropogenic biosphere. Ecological Monographs 85(3), 287-331.

\bibitem{Carballo2013}
Carballo, D.M. ed., 2012. {\it Cooperation and collective action: archaeological perspectives}. University Press of Colorado.

\bibitem{Roca2009}
Roca, C.P., Cuesta, J.A. and Sánchez, A., 2009. Evolutionary game theory: Temporal and spatial effects beyond replicator dynamics. Physics of life reviews 6, 208-249.

\bibitem{Helbing2009}
Helbing, D. and Yu, W., 2009. The outbreak of cooperation among success-driven individuals under noisy conditions. Proceedings of the National Academy of Sciences 106, 3680-3685.

\bibitem{Burtsev2006}
Burtsev, M. and Turchin, P., 2006. Evolution of cooperative strategies from first principles. Nature 440, 1041-1044.

\bibitem{McNally2012}
McNally, L., Brown, S.P. and Jackson, A.L., 2012. Cooperation and the evolution of intelligence. Proceedings of the Royal Society B 279, 3027-3034.

\bibitem{Sole2016}
Sol\'e, R., 2016. Synthetic transitions: towards a new synthesis. Philosophical Transactions of the Royal Society B 371, 20150438.




 

\end{references}



\end{document}